\documentclass[prb,preprint]{revtex4-1} 
\usepackage{amsmath}  
\usepackage{amsfonts} 
\usepackage{graphicx} 
\usepackage{braket}
\usepackage{nicefrac}

\usepackage{color}

\newcommand{\B}[1]{\mathbf{#1}}

\begin{document}
\title{A physically-motivated quantisation of the electromagnetic field}

\author{Robert Bennett}
\affiliation{The School of Physics and Astronomy, University of Leeds, Leeds LS2 9JT, United Kingdom}

\author{Thomas M. Barlow}
\affiliation{The School of Physics and Astronomy, University of Leeds, Leeds LS2 9JT, United Kingdom}

\author{Almut Beige}
\email{a.beige@leeds.ac.uk}
\affiliation{The School of Physics and Astronomy, University of Leeds, Leeds LS2 9JT, United Kingdom}

\date{\today}

\begin{abstract}
The notion that the electromagnetic field is quantised is usually inferred from observations such as the photoelectric effect and the black-body spectrum. However accounts of the quantisation of this field are usually mathematically motivated and begin by introducing a vector potential, followed by the imposition of a gauge that allows the manipulation of the solutions of Maxwell's equations into a form that is amenable for the machinery of canonical quantisation. By contrast, here we quantise the electromagnetic field in a less mathematically and more physically-motivated way. Starting from a direct description of what one sees in experiments, we show that the usual expressions of the electric and magnetic field observables follow from Heisenberg's equation of motion. In our treatment, there is no need to invoke the vector potential in a specific gauge and we avoid the commonly-used notion of a fictitious cavity that applies boundary conditions to the field. 
\end{abstract}

\maketitle 

\section{Introduction} 

As early as 1900, Planck introduced the idea of so-called basic {\em energy elements} into which the radiation of a blackbody could be divided. At first, this step was a mere mathematical trick which allowed him to derive a radiation law, which was consistent with experimental observations of the spectrum of a blackbody, from basic thermodynamical principles.\cite{Planck} Later on, Planck's work became recognised as the origin of quantum physics and his basic energy elements became known as {\em photons}. But this only happened later on, after Einstein put the reality of the energy quanta of the electromagnetic field on a more firm footing by linking it to the photoelectric effect.\cite{Einstein} Although it can be shown that the essence of the photoelectric effect does not require the quantisation of the radiation field,\cite{Lamb} the photon concept becomes unavoidable when describing the key behaviour of quantum light fields that are beyond the domain of the classical.\cite{Thorn,Knight}

Over the last decades, a wide range of experiments have been performed that test the properties of light at the quantum level. Most importantly, recent decades have seen the introduction and rapid improvement of devices with the ability to register single photons in the optical regime. Single photon detectors typically work by sensing an electrical signal that results from the absorption of a photon.\cite{detectors} Even single infrared photons can now be detected with an efficiency as high as $97\%$.\cite{detectors2} Moreover, single photon sources\cite{kuhn,Scheel2} are now an essential tool in many quantum optics laboratories worldwide. In addition to enabling novel technologies, like quantum cryptography\cite{BB84,Ekert,Gisin} and linear optics quantum computing,\cite{Knill,Kok} quantum optics experiments have helped us to answer a highly non-trivial but seemingly simple question,\cite{Nature} namely ``What is a photon?". 

When asked, most physicists now simply state that photons are the basic energy quanta of the electromagnetic field thus invoking the idea of the photon being a "particle" of light. This interpretation is not without its problems, a major one being that these energy quanta do not have a well-defined position, rather being infinite in extent. Others would avoid these problems by taking an instrumentalist approach where the photon is defined as simply whatever makes a detector click in an experiment sensitive enough to demonstrate the quantised nature of light. But this anti-realist interpretation is not particularly satisfying either. Field quantisation schemes in basic text books are often more mathematically than physically-motivated and therefore usually more detached from reality than is strictly necessary --- it could be argued that this adds unnecessary difficulties. 

The formal quantization of the electromagnetic field was first performed by Dirac in 1927.\cite{Dirac} Since then, most field quantisation schemes have relied on the mathematical fact that any function on a finite interval can be written as a Fourier series. More concretely, any real-valued function $f$ with argument $x \in (0,d)$ can be expanded in a series of exponentials,\cite{Math}  
\begin{eqnarray} \label{FS}
f(x) &=& \sum_{m=-\infty}^\infty c_m \, \exp \left({\rm i} m \, {2 \pi x \over d} \right) \, ,
\end{eqnarray}
where the $c_m$ are complex coefficients with $c_m = c_{-m}^*$. This is usually taken as the starting point when quantising the electromagnetic field inside a finite quantisation volume with certain electric field components vanishing at the boundaries.\cite{Fermi,Heitler,Abram,Cohen,Knoell,Glauber,Zuba,Barnett2,Loudonx,Knight,Drummond} Inspired by the above equation, the electromagnetic field observables are written as Fourier series of discrete sets of eigenfunctions which are the basic solutions of Maxwell's equations for the vector potential of the electromagnetic field in Coulomb gauge. The coefficients $c_m$ and $c_{-m}^*$ of these series are eventually replaced by photon annihilation and creation operators $\hat{c}_m$ and $\hat{c}_m^\dagger$, respectively. Subject to normalisation, the above-described canonical quantisation procedure automatically yields a harmonic oscillator Hamiltonian of the form 
\begin{eqnarray} \label{FS2}
H_{\rm field} &=& \sum_{m=1}^\infty \hbar \omega_m \, c_m^\dagger c_m + H_{\rm ZPE}
\end{eqnarray}
which sums over a discrete set of cavity frequencies $\omega_m$ and where $H_{\rm ZPE}$ denotes the energy of the vacuum, the so-called zero point energy. Afterwards, the infinite-volume limit is taken to yield the field observables of the free radiation field, thereby introducing a continuum of eigenfrequencies.

The purpose of this paper is to derive the field observables of the free radiation field from basic physical principles in a more direct way. The approach we present here is motivated by a recent experimental push towards integrated photonic devices for quantum information processing.\cite{rarity,fabio} Currently, a lot of effort is made worldwide to  combine linear optics elements, optical cavities and single photon sources\cite{kuhn} to realise a so-called quantum internet\cite{Kimble} and quantum networks for quantum simulations.\cite{Elica} When modelling such systems, it becomes important to use the same notion for the description of photons inside photonic devices as in linear optics scattering theory. One way of doing so is to extend the field quantisation scheme presented in this paper to the scattering of light through mirrors and optical resonators.\cite{Tom}

The physically-motivated field quantisation scheme which we present here has several advantages. For example, it does not invoke the solutions of Maxwell's equations in a specific gauge and there is no need to consider a finite quantisation volume with boundary conditions before being able to go to the infinite-volume limit. Instead, the starting point of our considerations is the experimental reality of what a photon is. We then notice that the basic principles of quantum physics for the construction of observables uniquely identify the relevant Hilbert space and the Hamiltonian $H_{\rm field}$ of the electromagnetic field inside a non-dispersive, non-absorbing, homogeneous medium. The usual expressions of the electric and magnetic field observables then follow from Heisenberg's equation of motion.

There are five sections in this paper. In Section \ref{Maxwellequation}, we review Maxwell's equations and discuss their basic solutions in the absence of any charges and currents. In Section \ref{1D}, we derive the corresponding observables of the quantised electromagnetic field for waves propagating along a particular axis from basic principles. In Section \ref{3D} we generalise these observables to the case of waves propagating in three dimensions. Finally, we summarise our findings in Section \ref{conclusions}.  

\section{Classical Electrodynamics} \label{Maxwellequation}

We begin by considering Maxwell's equations in a non-dispersive, non-absorbing, homogeneous medium with (absolute) permittivity $\varepsilon$ and permeability $\mu$. In the absence of any currents and charges, these equations are given by \cite{Teich}
\begin{align} \label{Maxwell}
 \nabla \cdot \B{E}(\B{r},t) &= 0, \, &
\nabla \times \B{E}(\B{r},t) &= - \frac{\partial \B{B}(\B{r},t)}{\partial t}  \, , \notag \\
 \nabla \cdot \B{B}(\B{r},t) &= 0, \, &
\nabla \times \B{B} (\B{r},t) &= \varepsilon \mu \, \frac{\partial \B{E}(\B{r},t)}{\partial t} \, . 
\end{align}
Here $\B{E}(\B{r},t)$ and $\B{B}(\B{r},t)$ are the electric and the magnetic field vector at time $t$ and at position $\B{r}$ within the medium, respectively. It is well known that the basic solutions of the above equations are travelling waves with wave vectors ${\bf k}= k \, \kappa$,
where $\kappa$ is a unit vector giving the relevant direction of the propagation.\cite{Teich} The frequency $\omega$ of these waves can assume any positive value and relates to the magnitude of the wave vector via the dispersion relation 
\begin{eqnarray} \label{positivek}
\omega &\equiv & \frac{k}{\sqrt{\varepsilon \mu}} \, .
\end{eqnarray}
In vacuum, this dispersion relation becomes  $\omega = k/\sqrt{\varepsilon_0\mu_0}\equiv ck $ where $\varepsilon_0$ and $\mu_0$ are the vacuum permittivity and permeability, respectively, and where $c$ is the speed of light. The field vectors $\B{E}$ and $\B{B}$ of these travelling waves are perpendicular to ${\bf k}$ and each other in order that the divergence of the electric and magnetic fields vanish as required by Maxwell's equations. 

\subsection{Propagation along the $x$ axis} \label{onedim}

\begin{figure}[b]
\center
\includegraphics[width=\columnwidth]{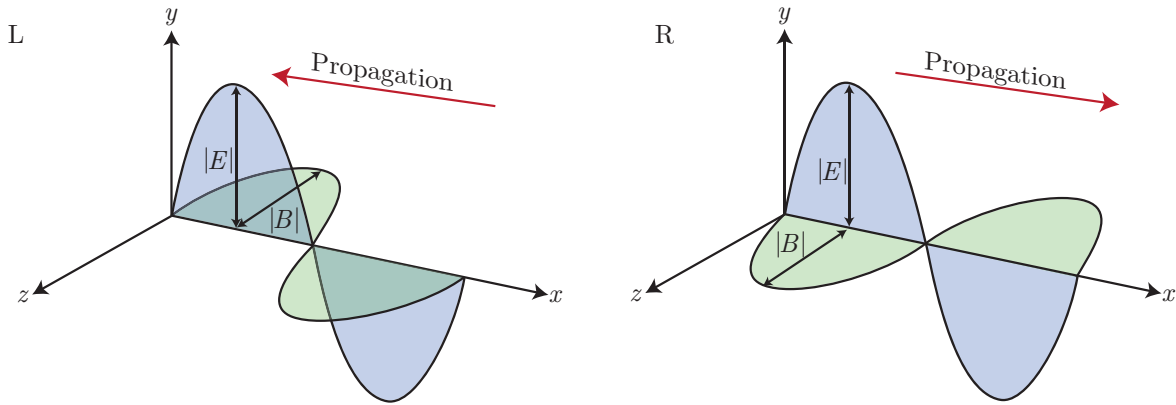} 
\caption{Schematic view of two travelling-wave solutions of Maxwell's equations propagating in one dimension at a fixed time $t$ and with a fixed polarisation and a particular frequency. The directions of the $\B{E}$ and $\B{B}$ fields are chosen as usual in classical electrodynamics, according to a right-hand rule for waves propagating to the right (R) and to the left (L) of the $x$ axis, respectively.} \label{WavesFig}
\end{figure}

Suppose a travelling wave propagates along the $x$ axis, and the corresponding $\B{E}$ and $\B{B}$ fields are aligned along the $y$ and $z$ axes, respectively. For a right-travelling wave, $\B{E}(\B{r},t)$ and $\B{B}(\B{r},t)$ can then be written as $\B{E}(\B{r},t)=(0,E(x,t),0)$ and $\B{B}(\B{r},t) = (0,0,B(x,t))$. Analogously, one can write $\B{E}(\B{r},t)=(0,E(x,t),0)$ and $\B{B}(\B{r},t) = (0,0,- B(x,t))$ for a left-travelling wave. Here $E(x,t)$ and $B(x,t)$ are defined as always having the same sign. Using this notation, the two Maxwell's equations in the right-hand column of Eq.~\eqref{Maxwell} become
\begin{eqnarray} \label{MW2}
\partial_x E (x,t) &=& \pm \partial_t B (x,t)  \, , \nonumber \\
\partial_x B (x,t) &=& \pm \varepsilon\mu \, \partial_t E (x,t) \, .
\end{eqnarray}
Which sign applies depends on whether the wave propagates in the positive or negative $x$ direction. In the following, we use the indices L and R to distinguish between left and right travelling waves. As illustrated in Fig.~\ref{WavesFig}, the plus sign in Eq.~\eqref{MW2} applies to the first (L) and the minus sign applies to the second (R) case. In one dimension, the total energy of the electromagnetic field equals 
\begin{equation} \label{classeng}
H_{\rm field} = A \int_{-\infty}^\infty dx \, {1 \over 2} \left[ \varepsilon E^2(x,t)+\frac{1}{\mu} B^2(x,t) \right] \, .
\end{equation}
Here $A$ is an area in the $y$-$z$ plane in which the Hamiltonian $H_{\rm field}$ is defined. This above expression is quadratic in both field components and constant in time. The latter can be shown using Eq.~(\ref{MW2}). 

\subsection{Propagation in three dimensions} \label{small}

In three dimensions, the basic solutions of Maxwell's equations are analogous to the case of waves propagating along the $x$ axis, but waves travelling in any direction of space and of any polarisation need to be considered. Each one of these travelling wave solutions is called a mode, and is characterised by a polarisation $\lambda$, which specifies the (positive) direction of its electric field, and a wave vector ${\bf k}$, which specifies its frequency and direction of propagation. The general solutions of Maxwell's equations are the superpositions of all possible modes $({\bf k}, \lambda)$.

\section{Field quantisation for propagation in one dimension} \label{1D}

As in the previous section, we consider a non-dispersive, non-absorbing, homogeneous medium with permittivity $\varepsilon$ and permeability $\mu$. We then begin our quantisation by returning to the question in the introduction: ``What is a photon?''.\cite{Nature} To answer this question, we point out that a detector that measures the energy of a very weak electromagnetic field produces discrete clicks. Single photon experiments have shown that these clicks are the signature of the fundamental property of the electromagnetic field, which is that the energy it carries is quantised.\cite{detectors,detectors2,kuhn,Scheel2,BB84,Ekert,Gisin,Knill,Kok} These energy quanta are called photons. 

\subsection{The relevant Hilbert space}

We know from observations that photons propagating in {\em one dimension} are characterised by their (positive) frequency, $\omega \in (0,\infty)$ and a direction of propagation, ${\rm X} = {\rm L}, {\rm R}$. In addition, they are characterised by their so-called polarisation, $\lambda =1,2$, which indicates the direction of their respective electric field vector.\cite{footnote} Combining these experimental facts with the rules of quantum physics, we then know that the Hilbert space for the description of the quantised electromagnetic field is spanned by tensor product states of the form
\begin{equation}  \label{StateDef}
\prod_{\omega =0}^\infty \prod_{{\rm X} = {\rm L}, {\rm R}} \prod_{\lambda = 1,2} \ket{n_{{\rm X}\lambda} (\omega)} \, , 
\end{equation}
where $n_{{\rm X}\lambda} (\omega)$ is the number of excitations in the $({\rm X},\lambda,\omega)$ photon mode. By construction, photons in different modes are in pairwise orthogonal states. In the following, we denote the ground state of the electromagnetic field, the so-called vacuum state, by $|0 \rangle$.  

\subsection{Field Hamiltonian}

The Hamiltonian of a system in the Schr\"odinger picture is its energy observable. Experiments have shown that the energy of an electromagnetic field increases by $\hbar \omega$, whenever a photon of frequency $\omega$ is added. Moreover we know that the energy eigenstates of the field are the states in Eq.~(\ref{StateDef}) with an integer number of photons in the field. Hence the Hamiltonian $\hat{H}_{\rm field}$ of the quantised electromagnetic field is such that
\begin{eqnarray} \label{StateDef2}
\hat{H}_{\rm field} \, \ket{n_{{\rm X}\lambda} (\omega)} &=& \big[ \hbar \omega \, n_{{\rm X}\lambda} (\omega) + H_{\rm ZPE} \big]  \ket{n_{{\rm X}\lambda} (\omega)} \, .
\end{eqnarray}
The constant $H_{\rm ZPE} $ in this equation denotes again the zero point energy, which we determine later on in this section. One way of obtaining an explicit expression for $H_{\rm field}$ is to sum over all the projectors onto the eigenstates of this operator multiplied by their respective eigenvalues, as it is usually done in quantum physics when constructing an observable.

Next we notice that the electromagnetic field has exactly the same energy level structure as a collection of independent harmonic oscillators with each of them characterised by a specific frequency $\omega$, a polarisation $\lambda$, and a direction of propagation X. This analogy suggests that it is possible to write the above field Hamiltonian in a much more compact form. To do so, we define the harmonic oscillator annihilation and creation operators $\hat a_{{\rm X}\lambda}(\omega)$ and $\hat a_{{\rm X}\lambda}^\dagger (\omega)$ such that 
\begin{eqnarray} \label{comm2}
\hat a_{{\rm X}\lambda} (\omega) \, \ket{n_{{\rm X}\lambda} (\omega)} &=& \sqrt{\smash[b]{n_{\rm{X}\lambda}(\omega)}}\, \ket{n_{{\rm X}\lambda} (\omega) - 1} \, , \notag \\
\hat a_{{\rm X}\lambda}^\dagger (\omega) \, \ket{n_{{\rm X}\lambda} (\omega)} &=&\sqrt{\smash[b]{n_{\rm{X}\lambda}(\omega)+1}} \, \ket{n_{{\rm X}\lambda} (\omega) + 1} \, .
\end{eqnarray}
These are photon annihilation and creation operators , i.e.~harmonic oscillator operators for each mode $(\omega, \lambda, X)$, and can be shown to obey the commutation relation
\begin{eqnarray} \label{comm}
\left[ \hat{a}_{{\rm X}\lambda} (\omega), \hat a_{{\rm X}'\lambda'}^\dagger (\omega') \right] &=& \delta_{{\rm X}{\rm X}'} \, \delta_{\lambda\lambda'} \, \delta (\omega - \omega') \, ,
\end{eqnarray}
since the states $\ket{n_{{\rm X}\lambda} (\omega)} $ form an orthonormal basis. Using the above notation, $\hat{H}_{\rm field}$ simplifies to 
\begin{eqnarray} \label{HFieldDefinition}
\hat{H}_{\rm field} &=&  \sum_{{\rm X} = {\rm L}, {\rm R}} \sum_{\lambda = 1,2} \int_0^\infty {\rm d} \omega \, \hbar \omega  \, \hat a_{{\rm X}\lambda}^\dagger (\omega) \hat a_{{\rm X}\lambda} (\omega) + H_{\rm ZPE} \, ,
\end{eqnarray}
where we sum over all possible photon modes $({\rm X},\lambda,\omega)$. One can easily check that the eigenstates and eigenvalues of the operator in Eq.~(\ref{HFieldDefinition}) are the same as the eigenstates and eigenvalues of $\hat{H}_{\rm field}$ in Eq.~(\ref{StateDef2}). 
 
\subsection{Electric and magnetic field observables}

We now seek expressions for the quantised electric and magnetic field observables $\hat{\B{E}} (x)$ and $\hat{\B{B}}(x) $ for waves propagating along the $x$ axis which correspond to the classical field amplitudes ${E} (x,t)$ and ${B} (x,t)$ in Section \ref{onedim}. To obtain these operators, we notice that the classical energy of the electromagnetic field is proportional to ${E} (x,t)^2$ and ${B} (x,t)^2$ (cf.~Eq.~(\ref{classeng})), while the field Hamiltonian $\hat{H}_{\rm field}$ is a quadratic function of the annihilation and creation operators $\hat{a}_{{\rm X} \lambda} (\omega)$ and $\hat{a}_{{\rm X} \lambda}^\dagger (\omega)$ (cf.~Eq.~(\ref{HFieldDefinition})). This suggests the following ansatz for the respective polarisation-dependent amplitudes of $\hat{\B{E}}(x)$ and $\hat{\B{B}}(x)$,
\begin{eqnarray} \label{sum100}
\hat{E}_\lambda(x) &=& \sum_{{\rm X} = {\rm L}, {\rm R}} \int_0^\infty {\rm d} \omega \,  f_{{\rm X}\lambda} (x,\omega) \, \hat a_{{\rm X}\lambda}(\omega) + {\rm H.c.} \, , \nonumber \\
\hat{B}_\lambda(x) &=& \sum_{{\rm X} = {\rm L}, {\rm R}} \int_0^\infty {\rm d} \omega \, g_{{\rm X}\lambda} (x,\omega) \, \hat a_{{\rm X}\lambda}(\omega) + {\rm H.c.}  \, ,
\end{eqnarray}
where $ f_{{\rm X}\lambda} (x,\omega)$ and $g_{{\rm X}\lambda} (x,\omega)$ are complex coefficients. Splitting the field observables in this way is well-justified, since both fields are linear and additive with respect to all possible photon modes $({\rm X},\lambda,\omega)$. 

We demand in the following that the expectation values of the observables of the quantised electromagnetic field behave as predicted by Maxwell's equations. Taking into account that the time derivative of an observable ${\cal \hat{O}}$ is given by Heisenberg's equation of motion yields
\begin{eqnarray} \label{partial}
\partial_t \langle {\cal \hat{O}}(x) \rangle &=& - {{\rm i} \over \hbar}  \left \langle \left[ {\cal \hat{O}}(x), \hat{H}_{\rm field} \right] \right \rangle 
\end{eqnarray}
with $\hat{H}_{\rm field}$ as in Eq.~(\ref{HFieldDefinition}). Hence we find consistency with Maxwell's equations as long as
\begin{eqnarray} \label{13}
\partial_x f_{{\rm X}\lambda} (x,\omega) &=& \mp {\rm i} \omega \, g_{{\rm X}\lambda} (x,\omega) \, , \nonumber \\
\partial_x g_{{\rm X}\lambda} (x,\omega) &=& \mp {\rm i} \varepsilon \mu \omega \, f_{{\rm X}\lambda} (x,\omega) \, .
\end{eqnarray}
The minus sign in this equation corresponds to ${\rm X} = {\rm L}$ and the plus sign corresponds to ${\rm X} = {\rm R}$ (cf.~Eq.~(\ref{MW2})). The general solution of this equation can be written as 
\begin{eqnarray} \label{effi2}
f_{{\rm X}\lambda} (x, \omega) &=& K_{\rm X,1} (\omega) \, {\rm e}^{- {\rm i} k x} + K_{\rm X,2} (\omega) \, {\rm e}^{{\rm i} k x} \, , \nonumber \\
g_{{\rm X}\lambda} (x, \omega) &=& \pm \sqrt{\varepsilon \mu} \left[ K_{\rm X,1} (\omega) \, {\rm e}^{- {\rm i} k x} - K_{\rm X,2} (\omega) \, {\rm e}^{{\rm i} k x} \right] 
\end{eqnarray}
with the always positive wave vector $k$ given in Eq.~(\ref{positivek}) and with the $K$ constants being complex functions of $\omega$ and X but independent of $x$, $t$ and $\lambda$. They can assume any value without contradicting Maxwell's equations. However, if we want the index ${\rm X}$ to specify the direction L or R, we need to ensure that the corresponding time-dependent expectation values $\langle \hat E (x) \rangle$ and $\langle \hat B (x) \rangle$ are either functions of $kx + \omega t$ or of $kx - \omega t$. This implies $K_{\rm L,2} = K_{\rm R,1} = 0$, while $K_{\rm L,1} $ and $K_{\rm R,2}$ remain unspecified.  

To determine $K_{\rm L,1} $ and $K_{\rm R,2}$ we now introduce a final constraint on the operators $\hat E (x)$ and $\hat B (x) $, which is that the expressions themselves must produce the quantum Hamiltonian \eqref{HFieldDefinition} in the previous section when substituted in the classical electromagnetic Hamiltonian \eqref{classeng}. In other words, we want that 
\begin{equation} \label{classeng2}
\hat H_{\rm field} = A \int_{-\infty}^\infty dx \, {1 \over 2} \left[ \varepsilon \hat{\B{E}}^2(x)+\frac{1}{\mu} \hat{\B{B}}^2(x) \right] 
\end{equation}
with $A$ defined as in Eq.~(\ref{classeng}). Combining this Hamiltonian with the above equations and performing the $x$ integration yields $\delta$-functions. Subsequently performing another integration, this finally results in 
\begin{equation} \label{class200}
\hat{H}_{\rm field} = 2 \pi \varepsilon A \sum_{\lambda = 1,2} \int_0^\infty {\rm d} \omega \left[ \left| K_{\rm L,1} \right|^2 \left( 2 \hat{a}_{{\rm L}\lambda}^\dagger (\omega) \hat{a}_{{\rm L}\lambda} (\omega) + 1 \right)
+ \left| K_{\rm R,2}\right|^2 \left( 2 \hat{a}_{{\rm R}\lambda}^\dagger (\omega) \hat{a}_{{\rm R}\lambda} (\omega) + 1 \right) \right] \, . 
\end{equation}
This operator becomes identical to the field Hamiltonian in Eq.~(\ref{HFieldDefinition}) when we choose the $K$ constants and the zero-point energy $H_{\rm ZPE } $ such that
\begin{eqnarray} \label{sum222}
\left| K_{\rm L,1}(\omega) \right|^2 = \left| K_{\rm R,2} (\omega)\right|^2 = {\hbar \omega \over 4 \pi \varepsilon A} 
~~ {\rm and} ~~ H_{\rm ZPE } =  A \int_0^\infty {\rm d} \omega \, {\textstyle {1 \over 2}} \hbar \omega \, .
\end{eqnarray}
After choosing phase factors (with no physical consequences) for the above constants $K_{\rm L,1}(\omega)$ and $K_{\rm R,2}(\omega)$ and substituting them into the above equations, we finally obtain the electric and magnetic field observables   
\begin{eqnarray} \label{sum301}
\hat{\bf E}(x) &=& {\rm i} \sum_{\lambda = 1,2} \int_0^\infty {\text d} \omega \, \sqrt{{\hbar \omega \over 4 \pi \varepsilon A}} \, {\rm e}^{- {\rm i} k x} \, \left[ \hat a_{{\rm L}\lambda} (\omega) - \hat a_{{\rm R}\lambda}^\dagger (\omega) \right] \B{e}_\lambda + {\text{H.c.}} \, , \nonumber \\
\hat{\bf B}(x) &=& - {{\rm i} }\sqrt{\varepsilon\mu} \sum_{\lambda = 1,2} \int_0^\infty {\text d} \omega \, \sqrt{{\hbar  \omega \over 4 \pi \varepsilon A}} \, {\rm e}^{- {\rm i} k x} \, \left[ \hat{a}_{{\text L}\lambda} (\omega) - \hat{a}_{\text{R}\lambda}^\dagger (\omega) \right] \left( \hat{\B{k}} \times \B{e}_\lambda \right) + {\text{\rm H.c.}} \, ,
\end{eqnarray}
with the (positive) wave number $k$ defined as in Eq.~(\ref{positivek}) and with $\hat{\B{k}} \equiv \B{k}/|\B{k}|$ being a unit vector in the $\B{k}$ direction. The vectors $\B{e}_1$ and $\B{e}_2$ are unit vectors orthogonal to $x$ and orthogonal to each other. For example, $\B{e}_1$ could be a vector oriented along the $y$ axis, while $\B{e}_2$ points in the direction of the $z$ axis. The above operators $\hat{\bf E}(x) $ and $\hat{\bf B}(x)$ are consistent with the usual textbook expressions for the quantised electromagnetic field propagating in one dimension.\cite{Loudonx,Knight,Drummond} 

\section{Field quantisation for propagation in three dimensions} \label{3D}

As pointed out in Section \ref{small}, the electromagnetic field for waves propagating in {\em three dimensions} has more degrees of freedom than in the case of waves propagating along a single axis. Otherwise, both have analogous properties. Taking this into account, we immediately see that the dimension of the relevant Hilbert space is significantly larger. Now photons traveling in all possible directions in a three-dimensional space have to be taken into account. To do so, we now introduce a set of annihilation and creation operators $\hat a_{{\bf k}\lambda}$ and $\hat a_{{\bf k}\lambda}^\dagger$ with the bosonic commutator relation
\begin{eqnarray} \label{comm3}
\left[ \hat{a}_{{\bf k}\lambda} , \hat a_{{\bf k}'\lambda'}^\dagger \right] &=& \delta ({\bf k} - {\bf k}') \, \delta_{\lambda,\lambda'} \, .
\end{eqnarray}
These photon operators are analogous to the one-dimensional operators $\hat a_{{\rm X}\lambda} (\omega)$ and $\hat a_{{\rm X}\lambda}^\dagger (\omega)$ in Eq.~(\ref{comm2}) but their respective direction X and their respective frequency $\omega$ are now specified by the direction of the wave vector ${\bf k}$ and a frequency $\omega_k$ defined as 
\begin{eqnarray}
\omega_k &\equiv & {|{\bf k}| \over \sqrt{\varepsilon \mu}} \, . 
\end{eqnarray}
Using the above notation and proceeding as in the previous section, the Hamiltonian of the quantised electromagnetic field in three dimensions can be written as
\begin{eqnarray} \label{HFieldDefinition3}
\hat{H}_{\rm field} &=& \sum_{\lambda = 1,2} \int {\text d}^3 \B{k} \, \hbar \omega_k  \, \hat a_{{\bf k}\lambda}^\dagger \hat a_{{\bf k}\lambda} + H_{\rm ZPE} \, ,
\end{eqnarray}
where $H_{\rm ZPE}$ denotes again the (infinite) energy of the vacuum. The Hilbert space of the electromagnetic field is the states space obtained when applying the above annihilation and creation operators onto this vacuum state. 

To obtain expressions for the electric and magnetic field observables $\hat{\B{E}}(\B{r})$ and $\hat{\B{B}}(\B{r})$ at a position $\B{r}$ within the field, we demand again that the expectation values of these operators evolve as predicted by Maxwell's equations. Imposing this condition for the travelling waves of any wave vector ${\bf k}$, this yields
\begin{eqnarray} \label{E&B}
\hat{\B{E}}(\B{r}) &=& \frac{\rm i}{(2\pi)^{3/2}} \sum_{\lambda = 1,2} \int {\rm d}^3 {\bf k} \, \sqrt{{\hbar \omega_k \over  2\varepsilon}} \, {\rm e}^{- {\rm i} \B{k}\cdot \B{r}} \, \hat{a}_{\B{k}\lambda} \, \B{e}_{\B{k}\lambda} + {\text{H.c.}} \, , \nonumber \\
\hat{\B{B}}(\B{r}) &=& - {{\rm i} \over (2\pi)^{3/2}} \sqrt{\varepsilon \mu} \sum_{\lambda = 1,2} \int {\rm d}^3 {\bf k} \, \sqrt{{\hbar \omega_k \over 2 \varepsilon}} \, {\rm e}^{- {\rm i} \B{k} \cdot \B{r}} \, \hat{a}_{\B{k}\lambda} \, \left( \hat{\B{k}} \times \B{e}_{\B{k}\lambda} \right) + {\text{\rm H.c.}} 
\end{eqnarray}
in analogy to Eq.~(\ref{sum301}). The vectors $\B{e}_{\B{k}\lambda}$ in these equations are polarisation vectors with $\B{e}_{\B{k}\lambda} \cdot \B{e}_{\B{k}\lambda'} = \delta_{\lambda,\lambda'}$ and ${\bf k} \cdot \B{e}_{\B{k}\lambda} = 0$. Allowing for negative wave numbers $k$, and not only positive ones, it is no longer necessary to distinguish left and right moving photons. The normalizing factors in the above equation are different from Eq.~(\ref{sum301}). However, one can easily check that a photon in the $({\bf k}, \lambda)$ mode has the energy $\hbar \omega_k$, when substituting the above field operators into the three-dimensional analog of Eq.~(\ref{classeng2}) with an infinite quantisation volume.  

For simplicity, this paper avoids a more rigorous derivation of Eq.~(\ref{E&B}) which would require a more detailed discussion of infinite-volume limits. Instead, the above derivation exploits the fact that the general form of Eq.~(\ref{E&B}), up to normalisation, is already well motivated by Eq.~(\ref{sum301}). Finally, we note that the above electric and magnetic field operators in Eq.~(\ref{E&B}) are again consistent with textbook expressions\cite{Loudonx,Knight,Drummond} but this time for the quantised electromagnetic field in three dimensions and in an infinite quantisation volume. 

\section{Conclusions} \label{conclusions}

In this paper we quantise the electromagnetic field in a more physically-grounded way than usually found in the literature.\cite{Dirac,Fermi,Heitler,Abram,Cohen,Knoell,Glauber,Zuba,Barnett2,Loudonx,Knight,Drummond}
Our approach might be criticised for being phenomenological instead of deriving equations via a rigorous canonical field quantisation method. However, any field quantisation scheme contains ad-hoc assumptions such as those made when extending the quantization volume to infinity. On top of this, the actual process of field quantization is usually based on the introduction of a vector potential of the classical electromagnetic field and a subsequent choice of a gauge which allows the manipulation of the solutions of Maxwell's equations into a form that is amenable for the process of canonical quantisation. While this process is completely mathematically justified, it can cause one to lose sight of the experimental reality. 

In this paper we took an alternative approach. Starting from a direct description of what one sees in experiments, we answer the question ``What is a photon?'' by pointing out that a photon is what causes a click at a detector.\cite{detectors,detectors2} Photons are the basic energy quanta of the electromagnetic field and can be characterised by their respective frequencies, directions, and polarisations. Doing so, we obtain the relevant field Hamiltonian $\hat H_{\rm field}$ in Eq.~(\ref{HFieldDefinition3}). Afterwards, we show that the usual expressions of the electric and magnetic field observables $\hat{\B{E}}(\B{r})$ and $\hat{\B{B}}(\B{r})$ in~Eq.~({\ref{E&B}) follow from Heisenberg's equation of motion. Our derivation of these observables does not invoke the introduction of a vector potential in an arbitrarily-chosen gauge. Furthermore, we do not need to consider a finite quantisation volume before being able to go to the infinite-volume limit. Our approach naturally lends itself to an infinite quantisation volume. However, extending the proposed field quantisation scheme to the electromagnetic field inside a two-sided optical resonator is relatively straightforward.\cite{Tom}

Finally we would like to point out that the derivation of field Hamiltonians which describe interactions, for example between a quantised electromagnetic field and an atom, require the consideration of observables which are canonically conjugate to the electric and the magnetic field observables. For this a choice of gauge must be made. To make this choice, experimental observations need to be taken into account, like the lack of photon emission from an atomic system in a free radiation field in the absence of external driving.\cite{Adam} In this way, it is guaranteed that the ground state of an interacting system minimises its free energy, as it should.

\begin{acknowledgments}
T.~B.~acknowledges financial support from a White Rose Studentship Network on Optimising Quantum Processes and Quantum Devices for future Digital Economy. R.~B.~thanks the UK Engineering and Physical Sciences Research Council for an EPSRC Doctoral Prize. Moreover, we thank Christopher C. Gerry and Stefan Weigert for stimulating discussions. 
\end{acknowledgments}


\begin{thebibliography}{10}
\bibitem{Planck}
M. Planck, {\em On the Law of the Energy Distribution in the Normal Spectrum}, Ann. Phys. {\bf 4}, 553 (1901).

\bibitem{Einstein}
A. Einstein, {\em On the quantum theory of radiation}, Physikalische Zeitschrift {\bf 18}, 121 (1917).

\bibitem{Lamb}
W. E. Lamb and M. O. Scully, {\em The photoelectric effect without photons}, in {\em Polarization, Matire et Rayonnement}, Volume in Honour of A. Kastler (Presses Universitaires de France, Paris, 1969).

\bibitem{Thorn}
J. J. Thorn, M. S. Neel, V. W. Donato, G. S. Bergreen, R. E. Davies, and M. Beck, Am. J. Phys. {\bf 72}, 1210 (2004).

\bibitem{Knight}
C. C. Gerry and P. L. Knight, {\em Introductory Quantum Optics}, Cambridge University Press (Cambridge, 2005).

\bibitem{detectors}
M. D. Eisaman, J. Fan, A. Migdall, and S. V. Polyakov, {\em Invited Review Article: Single-photon sources and detectors}, Rev. Sci. Instrum. {\bf 82}, 071101 (2011).

\bibitem{detectors2}
F. Marsili,	V. B. Verma, J. A. Stern, S. Harrington, A. E. Lita, T. Gerrits, I. Vayshenker, B. Baek, M. D. Shaw, R. P. Mirin, and S. W. Nam, {\em Detecting single infrared photons with 93\% system efficiency}, Nature Phot. {\bf 7}, 210 (2013).

\bibitem{kuhn}
A. Kuhn, M. Hennrich, and G. Rempe, {\em Deterministic Single-Photon Source for Distributed Quantum Networking}, 
Phys. Rev. Lett. {\bf 89}, 067901 (2002).

\bibitem{Scheel2}
S. Scheel, {\em Single-photon sources-an introduction}, J. Mod. Opt. {\bf 56}, 141 (2009).

\bibitem{BB84}
C. H. Bennett and G. Brassard, {\em Quantum cryptography: public key distribution and coin tossing},
Proceedings of IEEE International Conference on Computers, Systems and Signal Processing, p. 175 (1984).  

\bibitem{Ekert}
A. Ekert, {\em Quantum cryptography based on BellÕs theorem}, Phys. Rev. Lett. {\bf 67}, 661 (1991).

\bibitem{Gisin}
N. Gisin, G. G. Ribordy, W. Tittel, and H. Zbinden, {\em Quantum cryptography}, Rev. Mod. Phys. {\bf 74}, 145 (2002).

\bibitem{Knill}
E. Knill, R. Laflamme, and G. J. Milburn, {\em A scheme for efficient quantum computation with linear optics}, Nature {\bf  409} 46 (2001).

\bibitem{Kok}
P. Kok, W. J. Munro, K. Nemoto, T. C. Ralph, J. P. Dowling, and G. J. Milburn, {\em Linear optical quantum computing with photonic qubits}, Rev. Mod. Phys. {\bf 79}, 135 (2007).

\bibitem{footnote}
Notice, different orientations of the electric field vector $\B{E}$ can be realised by superposing two  travelling waves, one with $\lambda =1$ and one with $\lambda =2$.

\bibitem{Nature}
C. Roychoudhouri and R. Roy, {\em The Nature of Light: What is a photon?}, CRC Press, Taylor \& Francis Group (Boca Raton, 2008).

\bibitem{Dirac}
P. A.M. Dirac, {\em The Quantum Theory of the Emission and Absorption of Radiation}, Proc. R. Soc. London {\bf 114}, 243 (1927). 

\bibitem{Math}
G. B. Arfken and H. J. Weber, {\em Mathematical Methods for Physicists}, Academic Press (London, 2001).

\bibitem{Fermi}
E. Fermi, {\em Quantum Theory of Radiation}, Rev. Mod. Phys. {\bf 4}, 87 (1932). 

\bibitem{Heitler}
W. Heitler, {\em The Quantum Theory of Radiation}, Oxford University Press (Oxford, 1954). 

\bibitem{Abram}
I. Abram, {\em Quantum theory of light propagation: Linear medium}, Phys. Rev. A {\bf 35}, 4661 (1987).

\bibitem{Cohen}
C. Cohen-Tannoudji, J. Dupont-Roc, and G Grynberg, {\em Photons and Atoms: An Introduction to Quantum Electrodynamics}, John Wiley \& Sons, Inc. (New York, 1987).

\bibitem{Knoell}
L. Kn\"oll, W. Vogel, and D.-G. Welsch, {\em Action of passive, lossless optical systems in quantum optics}, Phys. Rev. A {\bf 36}, 3803 (1987).

\bibitem{Glauber}
R. J. Glauber and M. Lewenstein, {\em Quantum optics of dielectric media}, Phys. Rev. A {\bf 43}, 467 (1991).

\bibitem{Zuba}
M. O. Scully and M. S. Zubairy, {\em Quantum Optics}, Cambridge University Press (Cambridge, 1997).

\bibitem{Barnett2}
B. Huttner and S. M. Barnett, {\em Quantization of the electromagnetic field in dielectrics}, Phys. Rev. A {\bf 46}, 4306 (1992).

\bibitem{Loudonx}
R. Loudon, {\em The Quantum Theory of Light}, Oxford University Press (Oxford, 2000).

\bibitem{Drummond}
P. D. Drummond and M. Hilary, {\em The Quantum Theory of Nonlinear Optics}, Cambridge University Press (New York, 2014).

\bibitem{rarity}
A. Politi, M. J. Cryan, J. G. Rarity, S. Yu, and J. L. O'Brien, {\em Silica-on-Silicon Waveguide Quantum Circuits}, Science {\bf 320}, 646 (2008). 

\bibitem{fabio}
L. Sansoni, F. Sciarrino, G. Vallone, P. Mataloni, A. Crespi, R. Ramponi, and R. Osellame, Phys. Rev. Lett. {\bf 108}, 010502 (2012).

\bibitem{Kimble}
H. J. Kimble, {\em The quantum internet}, Nature 453, 1023 (2008).

\bibitem{Elica}
E. S. Kyoseva, A. Beige, and L. C. Kwek, {\em Coherent cavity networks with complete connectivity}, New J. Phys. {\bf 14}, 023023 (2012).

\bibitem{Tom}
T. M. Barlow, R. Bennett, and A. Beige, {\em A master equation for a two-sided optical cavity,} J. Mod. Opt. {\bf 62}, S11 (2015).

\bibitem{Teich}
J. D. Jackson, {\em Classical Electrodynamics}, John Wiley \& Sons, Inc. (New York, 1962).

\bibitem{Adam}
See for example A. Stokes, A. Kurcz, T. P. Spiller, and A. Beige, {\em Extending the validity range of quantum optical master equations}, Phys. Rev. A {\bf 85}, 053805 (2012) and references therein.
\end{thebibliography}
\end{document}